\documentclass[twocolumn,showpacs,superscriptaddress,preprintnumbers,amsmath,amssymb,aps,pre,10pt]{revtex4-1}

\usepackage{graphicx}% Include figure files
\usepackage{color}
\usepackage{dcolumn}% Align table columns on decimal point
\usepackage{bm}% bold math

\begin{document}

%\preprint{APS/123-QED}

\title{Influence of Cohesive Energy and Chain Stiffness on Polymer Glass Formation}

\author{Wen-Sheng Xu}
\email{wsxu@uchicago.edu}
\affiliation{James Franck Institute, The University of Chicago, Chicago, Illinois 60637, USA}

\author{Karl F. Freed}
\email{freed@uchicago.edu}
\affiliation{James Franck Institute, The University of Chicago, Chicago, Illinois 60637, USA}
\affiliation{Department of Chemistry, The University of Chicago, Chicago, Illinois 60637, USA}

\date{\today}% It is always \today, today, but any date may be explicitly specified

\begin{abstract}
The generalized entropy theory is applied to assess the joint influence of the microscopic cohesive energy and chain stiffness on glass formation in polymer melts using a minimal model containing a single bending energy and a single (monomer averaged) nearest neighbor van der Waals energy. The analysis focuses on the combined impact of the microscopic cohesive energy and chain stiffness on the magnitudes of the isobaric fragility parameter $m_P$ and the glass transition temperature $T_g$. The computations imply that polymers with rigid structures and weak nearest neighbor interactions are the most fragile, while $T_g$ becomes larger when the chains are stiffer and/or nearest neighbor interactions are stronger. Two simple fitting formulas summarize the computations describing the dependence of $m_P$ and $T_g$ on the microscopic cohesive and bending energies. The consideration of the combined influence of the microscopic cohesive and bending energies leads to the identification of some important design concepts, such as iso-fragility and iso-$T_g$ lines, where, for instance, iso-fragility lines are contours with constant $m_P$ but variable $T_g$. Several thermodynamic properties are found to remain invariant along the iso-fragility lines, while no special characteristics are detected along the iso-$T_g$ lines. Our analysis supports the widely held view that fragility provides more fundamental insight for the description of glass formation than $T_g$.
\end{abstract}

\pacs{64.70.pj, 83.80.Sg, 05.70.-a, 05.50.+q}

%\pacs{Valid PACS appear here}% PACS, the Physics and Astronomy Classification Scheme.
%\keywords{Suggested keywords}%Use showkeys class option if keyword display desired

\maketitle

\section{Introduction}

Although glasses are ubiquitous in nature and in our daily life, a deep microscopic understanding of the nature of the glass transition and the glassy state remains a fundamental challenge in condensed matter physics~\cite{RMP_83_587, JCP_137_080901, JCP_138_12A301, ARCMP_4_263}. Typically, the dynamics of the supercooled liquid slows down precipitously on approaching the glass transition temperature $T_g$, while structural changes in the liquid are rather mild. Most polymers readily form glasses upon cooling or compression, and hence, provide a unique opportunity for probing the physical mechanism of glass formation due to their distinctive molecular characteristics~\cite{Book_Floudas, RPP_68_1405, Mac_43_7875, JPCM_26_153101}. The fragility parameter $m$, quantifying the steepness of the temperature dependence of the dynamics, and $T_g$ constitute two of the most important quantities of polymer glass formation~\cite{JPCB_109_21285, JPCB_109_21350}, governing, for instance, whether a polymer material can be processed by extrusion, casting, ink jet, etc. Thus, the ability for the rational design of polymer materials with desired properties requires an understanding of the molecular factors influencing the magnitudes of $m$ and $T_g$.

The chemical structure of polymers has long been recognized as strongly influencing properties associated with glass formation, such as $m$ and $T_g$~\cite{Mac_26_6824}. For instance, polymers with rigid or sterically hindered backbones usually exhibit larger $m$ and greater $T_g$ than polymers with simple and less sterically hindered structures~\cite{Mac_26_6824}. This behavior reflects the strong impact of the backbone and side group structures on microscopic molecular properties, such as the cohesive energy and chain stiffness. Recent experimental data indicate that the microscopic cohesive energy and chain stiffness can even be modified significantly simply by altering the chemical species in the side groups~\cite{Mac_45_8430}. The shifts in molecular structure between different polymer materials inevitably lead to corresponding changes in the behavior of polymer glass formation. Despite substantial experimental and simulational evidence~\cite{Mac_26_6824, Mac_45_8430, Mac_31_4581, JCP_122_134505, SM_6_3430, Mac_41_7232,  PCCP_15_4604, PRL_97_045502, Mac_44_5044, Lan_29_12730, JCP_140_044901}, a predictive molecular theory that describes the impact of chemical structure on polymer glass formation has been slow to develop.

The generalized entropy theory (GET)~\cite{ACP_137_125} is a merger of the Adam-Gibbs (AG) relation between the structural relaxation time and the configurational entropy~\cite{JCP_43_139} and the lattice cluster theory (LCT) for the thermodynamics of semiflexible polymers~\cite{ACP_103_335}. Because the LCT enables probing the influence of various molecular details, such as the cohesive energy (described by the nearest neighbor van der Waals interaction energy $\epsilon$, and called the microscopic cohesive energy or just cohesive energy for short), the chain stiffness (described by the bending energy $E_b$), the molecular weights, and the monomer structures, on the thermodynamics of multicomponent polymer systems~\cite{ACP_103_335}, the GET has provided initial theoretical insights into the molecular origins of $m$ and $T_g$ by describing the sensitivity of $m$ and $T_g$ to separate, i.e., one-dimensional, variations of the cohesive energy, the chain stiffness, and the relative rigidity of the backbone and the side chains~\cite{ACP_137_125}. While the agreement with experiment of the non-trivial predictions from the GET provides strong validation of the theory, the goal of rational design of polymeric materials requires considering the additional complexities of real polymer materials. For instance, the previous calculations within the GET~\cite{JPCB_109_21285, JPCB_109_21350, ACP_137_125, JCP_123_111102, JCP_124_064901, JCP_125_144907, JCP_131_114905, ACR_44_194, JCP_138_234501} consider the simplest model in which all united atom groups (i.e., the basic units of the polymer chains) interact with a common monomer averaged interaction energy. However, different groups are known to have disparate, i.e., specific, interactions whose implications remain to be investigated within the LCT. Moreover, the variation in structure within a monomer implies that the monomer averaged interaction energy and the chain stiffness must all change simultaneously. While some experimental data of Sokolov and coworkers~\cite{Mac_45_8430, Mac_41_7232} demonstrate that $m$ and $T_g$ for different polymers can be understood within the GET by considering separate variations of properties with $\epsilon$ and $E_b$, other data exhibit quite perplexing behavior that probably arises due to the competitive influences of changes in the cohesive energy and chain stiffness between ``similar'' materials.

The present paper focuses on analyzing the nature of glass formation in polymer melts as described by the LCT with a model containing a single, monomer averaged interaction energy $\epsilon$ between all united atom groups in the chain. Specifically, the monomer averaged interaction model is used to explore the variation of $m$ and $T_g$ with $\epsilon$ and $E_b$. Although a previous work briefly illustrates the separate variation of $m$ and $T_g$ individually with $\epsilon$ and $E_b$~\cite{JCP_131_114905}, the present paper emphasizes the more complex combined influence of $\epsilon$ and $E_b$, thereby illuminating potential design concepts. Moreover, the results obtained here will be compared in a subsequent paper with a recently developed more realistic model where the LCT treats polymer melts with specific interactions~\cite{JCP_141_044909}. In particular, the chains in the more realistic model have the structure of poly($n$-$\alpha$-olefins) where the terminal segments on the side chains are assigned different, specific van der Waals interaction energies with other united atom groups. The greater realism introduced into the LCT and the GET by this new model enables testing the limits of validity of the present model with a single monomer averaged van der Waals energy~\cite{JPCB_109_21285, JPCB_109_21350, ACP_137_125, JCP_123_111102, JCP_124_064901, JCP_125_144907, JCP_131_114905, ACR_44_194, JCP_138_234501}. In addition, the subsequent work will study how the variation of the specific interactions can be used to exert greater control over the properties of designed materials. 

Section II provides some general background concerning the GET. Section III begins by delineating the general combined influence of $\epsilon$ and $E_b$ on the nature of polymer glass formation. Two algebraic functions $m(\epsilon, E_b)$ and $T_g(\epsilon, E_b)$ are found to recapture the computed dependence of $m$ and $T_g$ on $\epsilon$ and $E_b$. Section III then proceeds by introducing some important design concepts, such as iso-fragility and iso-$T_g$ lines. The iso-fragility lines are contours with constant $m$ but variable $T_g$ and other properties. Our analysis demonstrates that many properties, e.g., the entropy, the polymer volume fraction and the relaxation times at characteristic temperatures, such as $T_g$, remain invariant along the iso-fragility lines. By contrast, no special characteristics are found along the iso-$T_g$ lines. Our results support the widespread view that the concept of fragility provides more fundamental insight into glass formation than the glass transition temperature $T_g$.

\section{Polymer glass formation within the generalized entropy theory}

The configurational entropy plays a central role in the GET~\cite{ACP_137_125} as in the classic entropy theories of glass formation by Gibbs and DiMarzio (GD)~\cite{JCP_28_373} and by Adam and Gibbs (AG)~\cite{JCP_43_139}. These theoretical approaches build upon the well-known fact that the rapid increase in the viscosity and structural relaxation time on cooling towards the glass transition temperature is accompanied by a precipitous drop in the fluid entropy~\cite{Nature_410_1924}. The GET~\cite{ACP_137_125} merges the LCT for the thermodynamics of semiflexible polymers with the AG relation between the structural relaxation time and the configurational entropy. Hence, the GET permits computing characteristic temperatures and fragility and addressing the influence of various molecular characteristics, such as monomer structure, on polymer glass formation. Therefore, the GET involves a significant extension of the scope beyond that of GD theory~\cite{JCP_28_373}, which effectively only focuses on the ``ideal'' glass transition temperature where the configurational entropy extrapolates to zero.

The LCT yields an analytical expression for the specific Helmholtz free energy $f$ (i.e., the total Helmholtz free energy per lattice site) of a semiflexible polymer melt~\cite{ACP_103_335}, as a function of polymer volume fraction $\phi$, temperature $T$, cohesive energy $\epsilon$, bending energy $E_b$ as well as molecular weight $M$ and a set of geometrical indices that reflect the size, shape and bonding patterns of monomers. The explicit expression for $f$ can be obtained in ref~\citenum{ACP_103_335} with some corrections given in ref~\citenum{JCP_141_044909}. We briefly explain the physical meaning of the key molecular parameters $\epsilon$ and $E_b$ in the LCT. The microscopic cohesive energy $\epsilon$ enters the LCT in terms of the Mayer $f$-function, which, in turn, is treated using a high temperature expansion~\cite{ACP_103_335, Mac_24_5076}, and a convention that a positive $\epsilon$ describes the net attractive van der Waals interactions between nearest neighbor united atom groups. The bending energy $E_b$ represents the conformational energy difference between, e.g., \textit{trans} and \textit{gauche} conformations for a pair of consecutive bonds~\cite{ACP_103_335}. The \textit{trans} conformation corresponds to consecutive parallel bonds and is ascribed a vanishing bending energy, while $E_b$ is prescribed to a \textit{gauche} pair of sequential bonds lying along orthogonal directions. Chains are fully flexible for $E_b=0$, while they become completely rigid in the limit $E_b\rightarrow\infty$. Also, side chains with two or more united atom groups may have a separate bending energy~\cite{JCP_124_064901}.

Within the GET, polymer glass formation is treated as a broad transition with four characteristic temperatures whose determination is accomplished by first computing the LCT configurational entropy density $s_c$~\cite{JCP_119_5730}, i.e., the configurational entropy per lattice site. This $s_c$ exhibits a maximum as the temperature $T$ varies at constant pressure~\cite{ACP_137_125}, an essential feature for use in the AG model. Recent computations~\cite{PC1} also indicate that the LCT configurational entropy density $s_c$ is almost identical to the ordinary entropy density $s=-\partial f/\partial T |_{\phi}$ at the same thermodynamic conditions because the lattice model is essentially devoid of vibrational contributions. Since the ordinary entropy density very closely approximates the configurational entropy density and since the former is much easier to calculate in the LCT, the calculations employ the ordinary entropy density in the present work. For simplicity, the ordinary entropy density is also called the entropy density in the following. Also, all the calculations are performed at constant pressure $P$, defined by
\begin{eqnarray}
P=-\left.\frac{\partial F}{\partial V}\right|_{N_p,T}=-\left.\frac{1}{V_{\text{cell}}}\frac{\partial F}{\partial N_l}\right|_{N_p,T},
\end{eqnarray}
where $V$ is the volume of the system, $N_p$ is the number of polymer chains, and $V_{\text{cell}}=a_{\text{cell}}^3$ is the volume associated with a single lattice site that is set to be $a_{\text{cell}}=2.7$\AA{} in all the calculations.

\begin{figure}
 \centering
 \includegraphics[angle=0,width=0.45\textwidth]{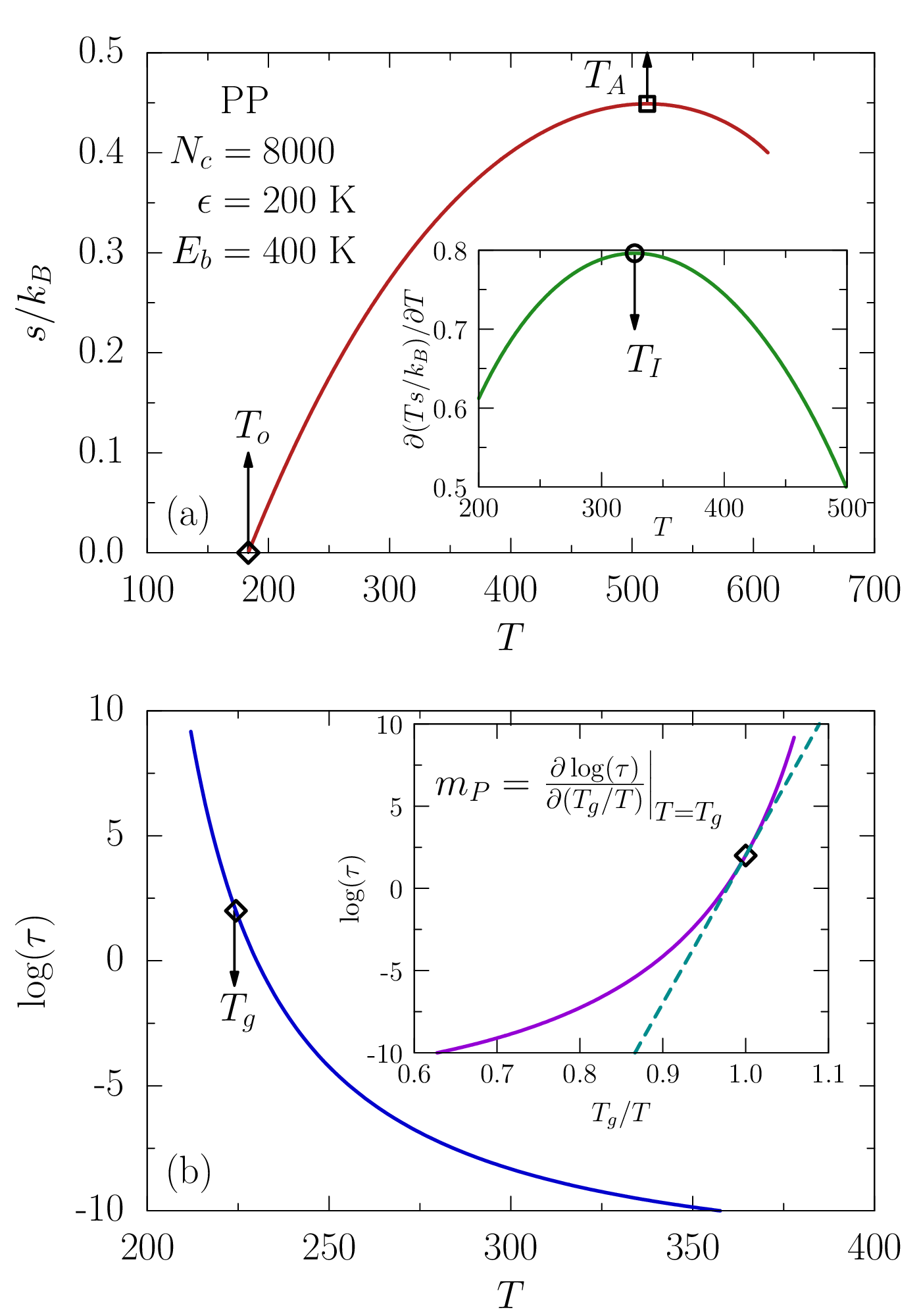}
 \caption{Illustration of the determination of various characteristic temperatures and the fragility parameter within the generalized entropy theory (GET). (a) Temperature $T$ dependence of the entropy density $s/k_B$ calculated from the lattice cluster theory (LCT)  for a melt of chains possessing the structure of poly(propylene) (PP) with polymerization index $N_c=8000$, cohesive energy $\epsilon=200$ K and bending energy $E_b=400$ K, at a constant pressure of $P=1$ atm. Three characteristic temperatures are well defined in the curve of $s(T)/k_B$: the ideal glass transition temperature $T_o$ where the entropy density extrapolates to zero, the onset temperature $T_A$ where the entropy density displays a maximum, and the crossover temperature $T_I$ which corresponds to the inflection point in the curve of $Ts(T)/k_B$. Equivalently, $T_I$ can be computed by finding the maximum of $\partial (Ts/k_B)/\partial T$, as shown in the inset to Figure 1a. (b) Structural relaxation time $\tau$ for the same melt as a function of $T$, calculated by combining the LCT with the Adam-Gibbs (AG) relation. The glass transition temperature $T_g$ is identified by the common empirical definition $\tau(T_g)=100$ s. The inset to Figure 1b illustrates the determination of the isobaric fragility parameter $m_P$ from an Angell plot~\cite{Science_267_1924}.}
\end{figure}

The LCT computations for the temperature dependence of the entropy density $s(T)$ enable the direct determination of three characteristic temperatures of glass formation, namely, the ``ideal'' glass transition temperature $T_o$,  the onset temperature $T_A$, and the crossover temperature $T_I$ (Figure 1a). $T_o$ corresponds to the temperature where $s$ extrapolates to zero as in the GD theory, $T_A$ signals the onset of non-Arrhenius behavior of the relaxation time and is found from the maximum in $s(T)$. The crossover temperature $T_I$ separates two temperature regimes with qualitatively different dependences of the relaxation time on temperature and is evaluated from the inflection point in $Ts(T)$. Equivalently, $T_I$ can be determined by finding the maximum of $\partial (Ts)/\partial T$, as shown in the inset to Figure 1a. Although the fourth characteristic temperature, i.e., the glass transition temperature $T_g$, might be obtained from a Lindermann criterion~\cite{JPCB_109_21285}, its conventional definition requires knowledge of the temperature dependence of the relaxation time $\tau$. To this end, the GET invokes the AG relation~\cite{JCP_43_139},
\begin{equation}
\tau=\tau_\infty\exp[\beta\Delta\mu s^\ast/s(T)],
\end{equation}
where $\tau_\infty$ is the high temperature limit of the relaxation time, $\beta=1/(k_BT)$ with $k_B$ being Boltzmann's constant, $\Delta\mu$ is the limiting temperature independent activation energy at high temperatures, and $s^\ast$ is the high temperature limit of $s(T)$ (i.e., the maximum of the entropy density calculated from the LCT). $\tau_\infty$ is set to be $10^{-13}$ s in the GET, which is a typical value for polymers~\cite{PRE_67_031507}. Motivated by the experimental data~\cite{PRE_67_031507}, the GET estimates the high temperature activation energy from the empirical relation $\Delta\mu=6k_BT_I$; more discussion of this empirical relation appears in ref~\citenum{ACP_137_125}. Thus, the relaxation time is computed within the GET without adjustable parameters beyond those used in the LCT for the thermodynamics of semiflexible polymers. The GET then identifies $T_g$ using the common empirical definition $\tau(T_g)=100$ s (Figure 1b).

Once the temperature dependence of the relaxation time is known, other related quantities can be also calculated from the GET. For instance, the fragility parameter, which quantifies the steepness of the temperature dependence of the relaxation time, can be determined at constant pressure, e.g., from the standard definition,
\begin{equation}
m_P=\left.\frac{\partial \log (\tau)}{\partial (T_g/T)}\right|_{P, T=T_g},
\end{equation}
where $m_P$ denotes the isobaric fragility parameter. The inset to Figure 1b illustrates the determination of $m_P$ from an Angell plot~\cite{Science_267_1924}.

\section{Results and discussion}

This section begins by discussing how the properties associated with glass formation vary under the combined influence of the microscopic cohesive $\epsilon$ and bending $E_b$ energies in polymer melts with monomer averaged interactions. This discussion then naturally leads to the introduction of the concepts of iso-fragility and iso-$T_g$ lines and the exploration of melt properties along these lines.

\subsection{Combined influence of microscopic cohesive and bending energies on polymer glass formation}

A previous paper~\cite{JCP_131_114905} investigates the variation of fragility and $T_g$ individually with $\epsilon$ and with $E_b$. However, the combined variation with the cohesive and bending energies yields a more detailed behavior. The calculations consider chains with the structure of PP because this choice requires the minimal number of parameters in the LCT. The simplest model for poly($n$-$\alpha$-olefins) with longer side groups $n>1$ contain separate bending energies for the backbone and side groups, thus adding another parameter~\cite{JCP_124_064901}. The subsequent figures all use a common parameter set: the lattice coordination number is $z=6$; the pressure is $P=1$ atm; the cell volume parameter is $a_{\text{cell}}=2.7$\AA{}; and the polymerization index is chosen to be $N_c=8000$, corresponding to a polymer melt of chains with high molecular weight.

\begin{figure}
 \centering
 \includegraphics[angle=0,width=0.45\textwidth]{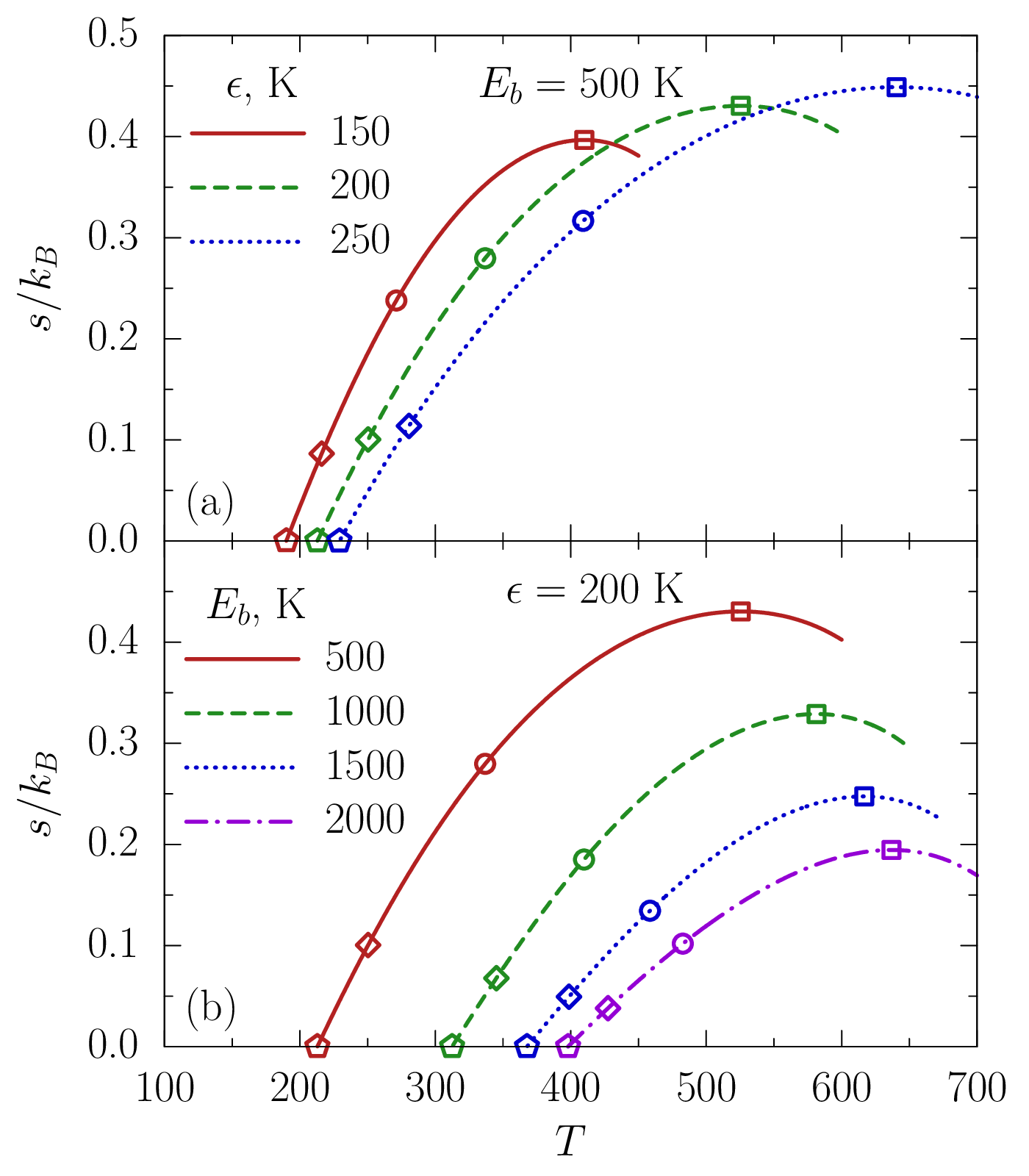}
 \caption{(a) The entropy density $s/k_B$ as a function of $T$ for various cohesive energies $\epsilon$ but fixed bending energy $E_b$. (b) The entropy density $s/k_B$ as a function of $T$ for various bending energies $E_b$ but fixed cohesive energy $\epsilon$. Squares, circles, diamonds and pentagons designate the positions of characteristic temperatures $T_A$, $T_I$, $T_g$ and $T_o$, respectively.}
\end{figure}

Figure 2 displays the entropy density for various cohesive energies with constant bending energy and for various bending energies with constant cohesive energy. As expected, both cohesive energy and bending energy strongly affect the entropy density. The curves for the entropy density in Figure 2 shift to higher temperatures as either $\epsilon$ or $E_b$ grows. Consequently, all characteristic temperatures elevate upon increasing either $\epsilon$ or $E_b$. Moreover, the magnitude of the entropy density at each characteristic temperature $s_{T_{\alpha}}$ ($T_{\alpha}=T_A, T_I$ or $T_g$) is altered in opposite directions as either $\epsilon$ or $E_b$ increases. Specifically, $s_{T_{\alpha}}$ grows with $\epsilon$ but drops with $E_b$. Furthermore, the characteristic temperature ratio $T_A/T_o$ grows with increasing $\epsilon$ or decreasing $E_b$, implying that the breadth of the glass-formation process can be controlled by adjusting the cohesive energy, the chain stiffness, or both. The latter observation is important because the breadth of glass formation is often suggested as being governed by the fragility~\cite{PRE_67_031507, JNCS_202_164, JNCS_223_207}.

\begin{figure}
 \centering
 \includegraphics[angle=0,width=0.45\textwidth]{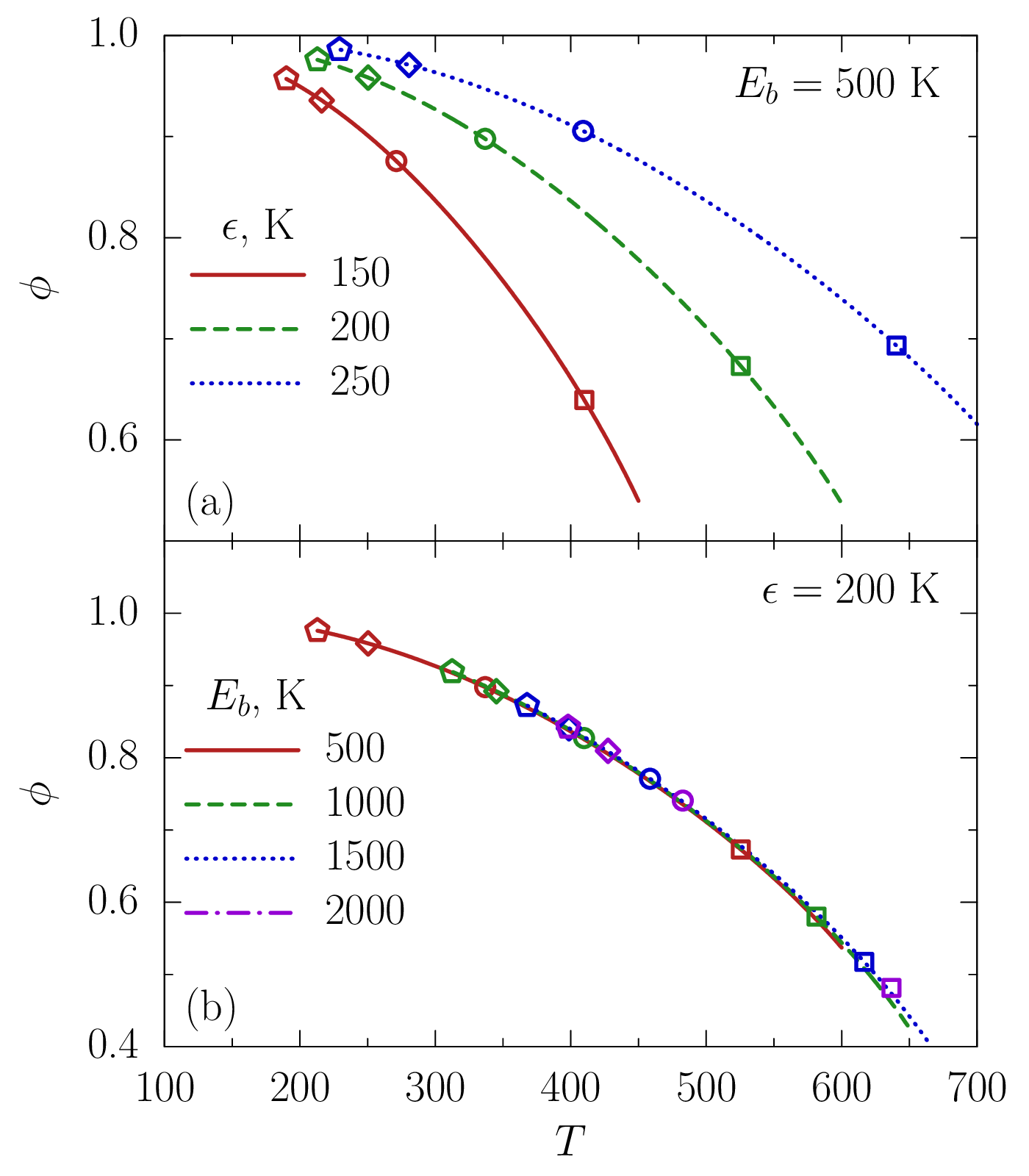}
 \caption{(a) The polymer volume fraction $\phi$ as a function of $T$ for various cohesive energies $\epsilon$ but fixed bending energy $E_b$. (b) The polymer volume fraction $\phi$ as a function of $T$ for various bending energies $E_b$ but fixed cohesive energy $\epsilon$. Squares, circles, diamonds and pentagons designate the positions of characteristic temperatures $T_A$, $T_I$, $T_g$ and $T_o$, respectively.}
\end{figure}

Knowledge of the equation of state (EOS) is also crucial for understanding, for instance, the pressure dependence of glass formation~\cite{JCP_138_234501}. In addition, the EOS provides complementary information to the entropy density and therefore aids in understanding the joint influence of cohesive and bending energies on polymer glass formation. Hence, we additionally explore how the temperature dependence of the polymer volume fraction $\phi$ is affected by variations of the microscopic cohesive and bending energies at constant pressure. The EOS data in Figure 3 clearly indicate that the temperature dependence of $\phi$ is dominated by the cohesive energy $\epsilon$. This anticipated result arises since $\epsilon$ provides the measure of the net attractive interactions between united atom groups in the LCT and thus a larger $\epsilon$ is expected to produce a denser system as evidenced in Figure 3a. Hence, polymers pack more efficiently as the cohesive energy increases. Consequently, the volume fraction at each characteristic temperature $\phi_{T_{\alpha}}$ ($T_{\alpha}=T_A, T_I, T_g$ or $T_o$) increases with $\epsilon$. The bending energy $E_b$, on the other hand, has almost no effect on the temperature dependence of $\phi$ (Figure 3b). However, $\phi_{T_{\alpha}}$ drops significantly as $E_b$ elevates, a trend that agrees with the general observations that more free volume exists in the glassy state of more rigid polymers.

\begin{figure}
 \centering
 \includegraphics[angle=0,width=0.45\textwidth]{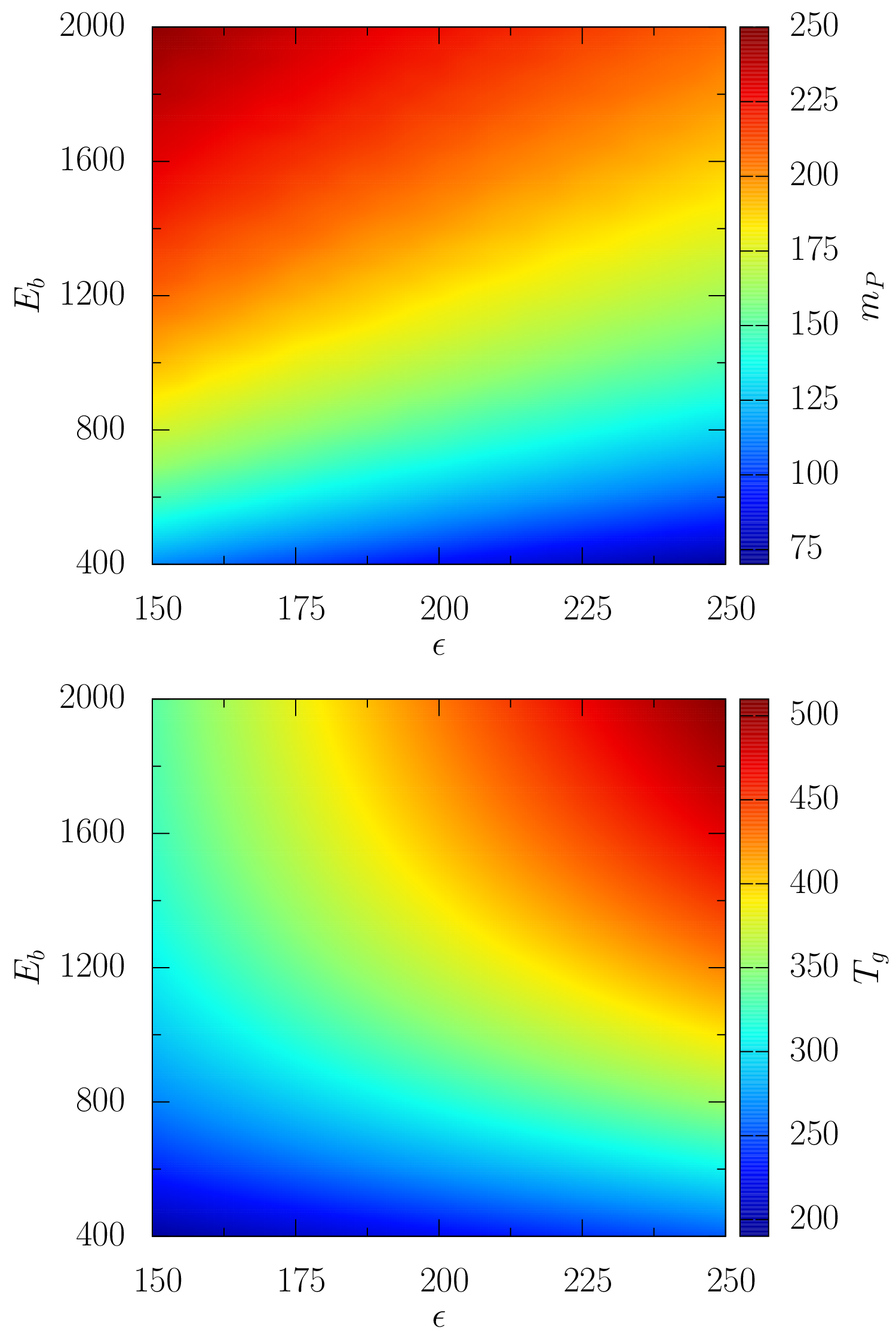}
 \caption{Contour plots of the isobaric fragility parameter $m_P$ (upper panel) and the glass transition temperature $T_g$ (lower panel) in the plane of cohesive energy $\epsilon$ and bending energy $E_b$.}
\end{figure}

We now discuss the combined effects of $\epsilon$ and $E_b$ on both $m_P$ and $T_g$. In line with previous calculations,~\cite{JCP_131_114905} the contour plots in Figure 4 clearly indicate that increasing $\epsilon$ and $E_b$ produces opposite shifts in $m_P$, while $T_g$ is altered in the same direction. The trend of an increasing $m_P$ or $T_g$ with $E_b$ is of course present over a limited range because $m_P$ and $T_g$ must saturate when the chains become very stiff, i.e., $E_b$ is sufficiently large (see Figure S1 in the Supporting Information). Two simple algebraic equations fairly accurately capture the computed combined variations of $m_P$ and $T_g$ with $\epsilon$ and $E_b$,
\begin{equation}
m_P=\frac{a_0+a_1\epsilon+a_2\epsilon^2+(b_0+b_1\epsilon)E_b}{1+(c_0+c_1\epsilon+c_2\epsilon^2)E_b}, 
\end{equation}
\begin{equation}
T_g=\frac{u_0+u_1/\epsilon+u_2/\epsilon^2+(v_0+v_1/\epsilon+v_2/\epsilon^2)E_b}{1+(w_0+w_1/\epsilon+w_2/\epsilon^2)E_b}, 
\end{equation}
where the fitted parameters $a_{\alpha}(\alpha=0,...,2)$, $b_{\alpha}(\alpha=0,1)$, $c_{\alpha}(\alpha=0,...,2)$, $u_{\alpha}(\alpha=0,...,2)$, $v_{\alpha}(\alpha=0,...,2)$ and $w_{\alpha}(\alpha=0,...,2)$ can be found in the caption of Figure S1 in the Supporting Information. Other forms of the equation may satisfactorily describe the data as well, but eqs 4 and 5 have been chosen to assure the observed saturation of $m_P$ and $T_g$ for large $E_b$.

\begin{figure}
 \centering
 \includegraphics[angle=0,width=0.45\textwidth]{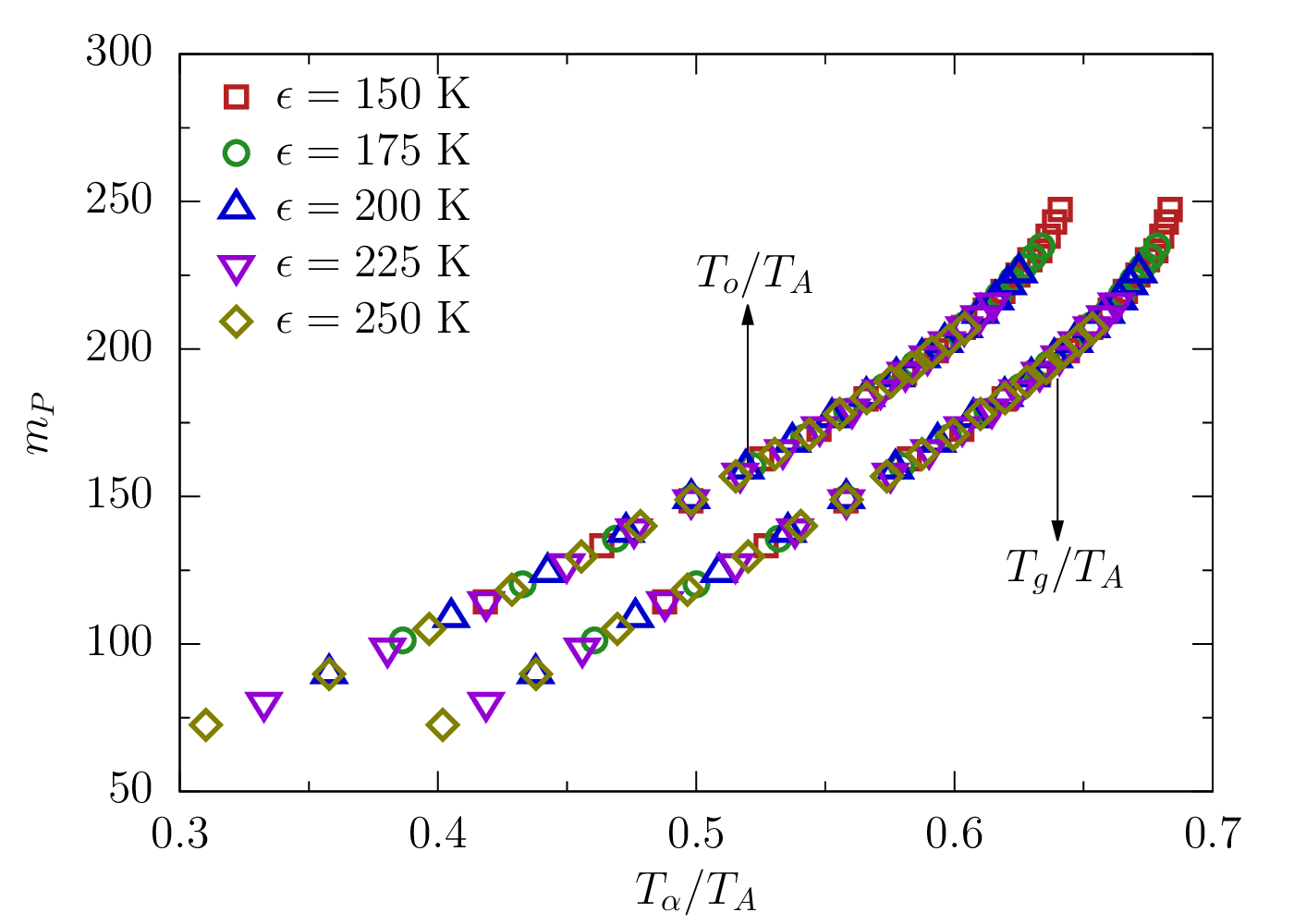}
 \caption{Correlation between isobaric fragility parameter $m_P$ and different ratios of characteristic temperature $T_{\alpha}/T_A$. The bending energy $E_b$ varies from $400$ K to $2000$ K for each $\epsilon$.}
\end{figure}

Figure 5 further demonstrates that a master curve exists between the isobaric fragility parameter $m_P$ and the characteristic temperature ratios, implying that the commonly used $m_P$ indeed correlates with ratios of the characteristic temperatures in some cases. The idea that the breadth of glass formation is related to fragility is certainly not new, but the plot in Figure 5 establishing its universality is presented here for the first time. Our results support the contention that the breadth of the glass-formation process provides a promising measure for the fragility of glass-forming liquids~\cite{ACP_137_125}.

\subsection{Iso-fragility and iso-$T_g$ lines}

The strong influence of the cohesive and bending energies on polymer glass formation clearly demonstrates that the fragility and the glass transition temperature can be finely tailored by adjusting these molecular parameters. In particular, we identify two types of special lines in the $\epsilon$-$E_b$ plane, along which either the fragility parameter $m_P$ or the glass transition temperature $T_g$ remains constant. These lines are naturally called iso-fragility and iso-$T_g$ lines, respectively, and their existence is indeed quite apparent in the contour plots of Figure 4. An exploration of various properties along the iso-fragility and iso-$T_g$ lines provides better understanding of the physical significance of $m_P$ and $T_g$ in glass-forming polymers, as discussed below.

\begin{figure}
 \centering
 \includegraphics[angle=0,width=0.45\textwidth]{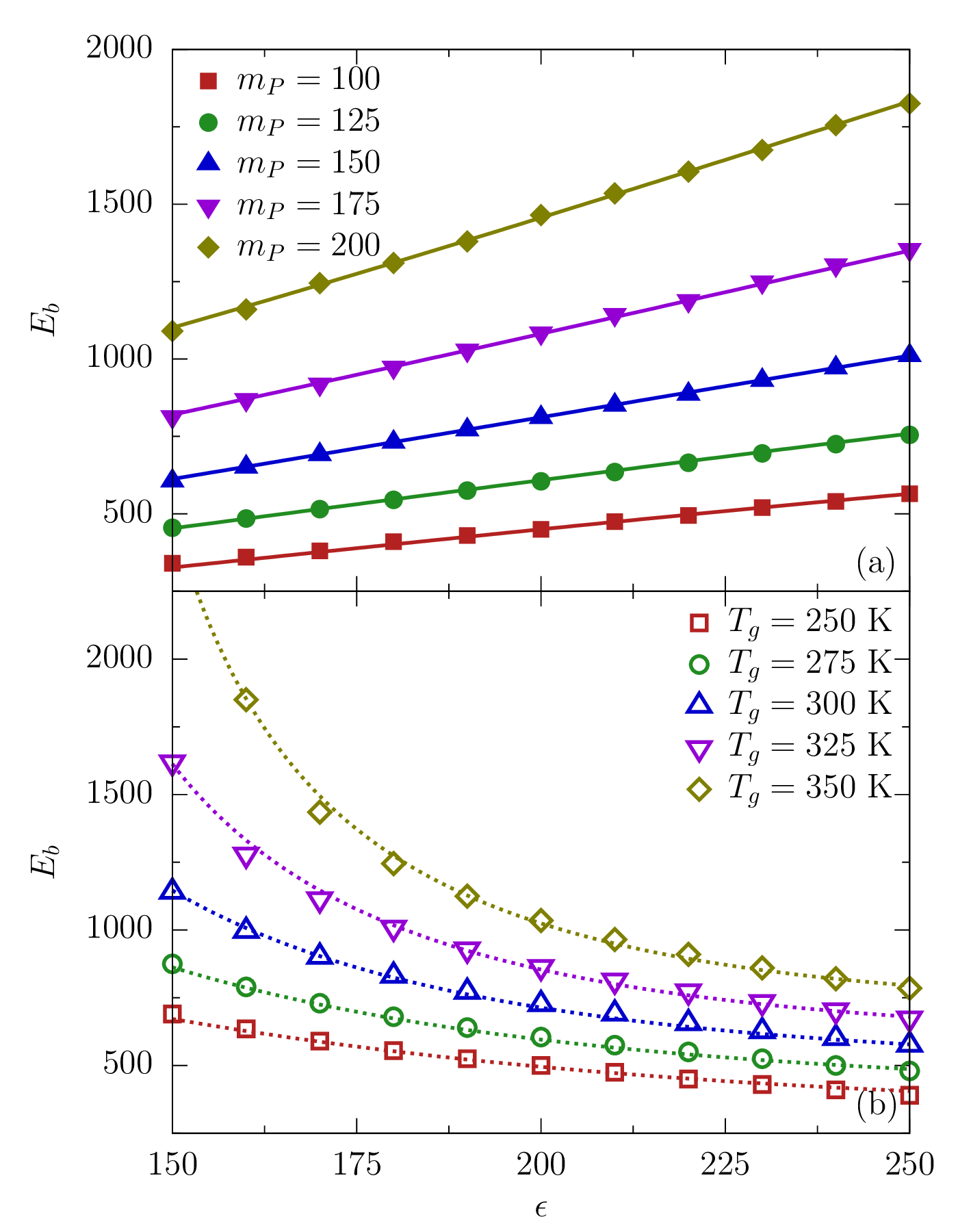}
 \caption{(a) Iso-fragility lines and (b) iso-$T_g$ lines in the plane of cohesive energy $\epsilon$ and bending energy $E_b$ for several representative values of $m_P$ and $T_g$. Solid lines in (a) and dotted lines in (b) are the results of eqs 4 and 5 with the fitting parameters given in the caption of Figure S1 in the Supporting Information.}
\end{figure}

Several iso-fragility and iso-$T_g$ lines are displayed in Figure 6 as curves of $\epsilon$ vs. $E_b$ for representative values of $m_P$ and $T_g$, respectively. The lines in Figure 6 display approximations to the iso-fragility and iso-$T_g$ lines obtained from eqs 4 and 5 with the fitting parameters given in the caption of Figure S1 in the Supporting Information. The fitted expressions clearly agree well with the calculations along the iso-fragility and iso-$T_g$ lines. Solutions for $E_b$ fail to exist for $T_g=350$ K on the iso-$T_g$ line with $\epsilon=150$ K because $T_g$ saturates at a smaller limit than $350$ K for $\epsilon=150$ K (see Figure S1 in the Supporting Information).

Figure 6a indicates that $E_b$ grows approximately linearly with $\epsilon$ along an iso-fragility line and that the slope grows with $m_P$.  Figure 6b exhibits the trend that $E_b$ decreases with $\epsilon$ along the iso-$T_g$ lines, so chains must become more flexible for larger cohesive energies in order for the system to achieve the same $T_g$ as at a smaller $\epsilon$. The drop in $E_b$ with $\epsilon$ along the iso-$T_g$ lines is quicker for low than high $\epsilon$.

\begin{figure}
 \centering
 \includegraphics[angle=0,width=0.45\textwidth]{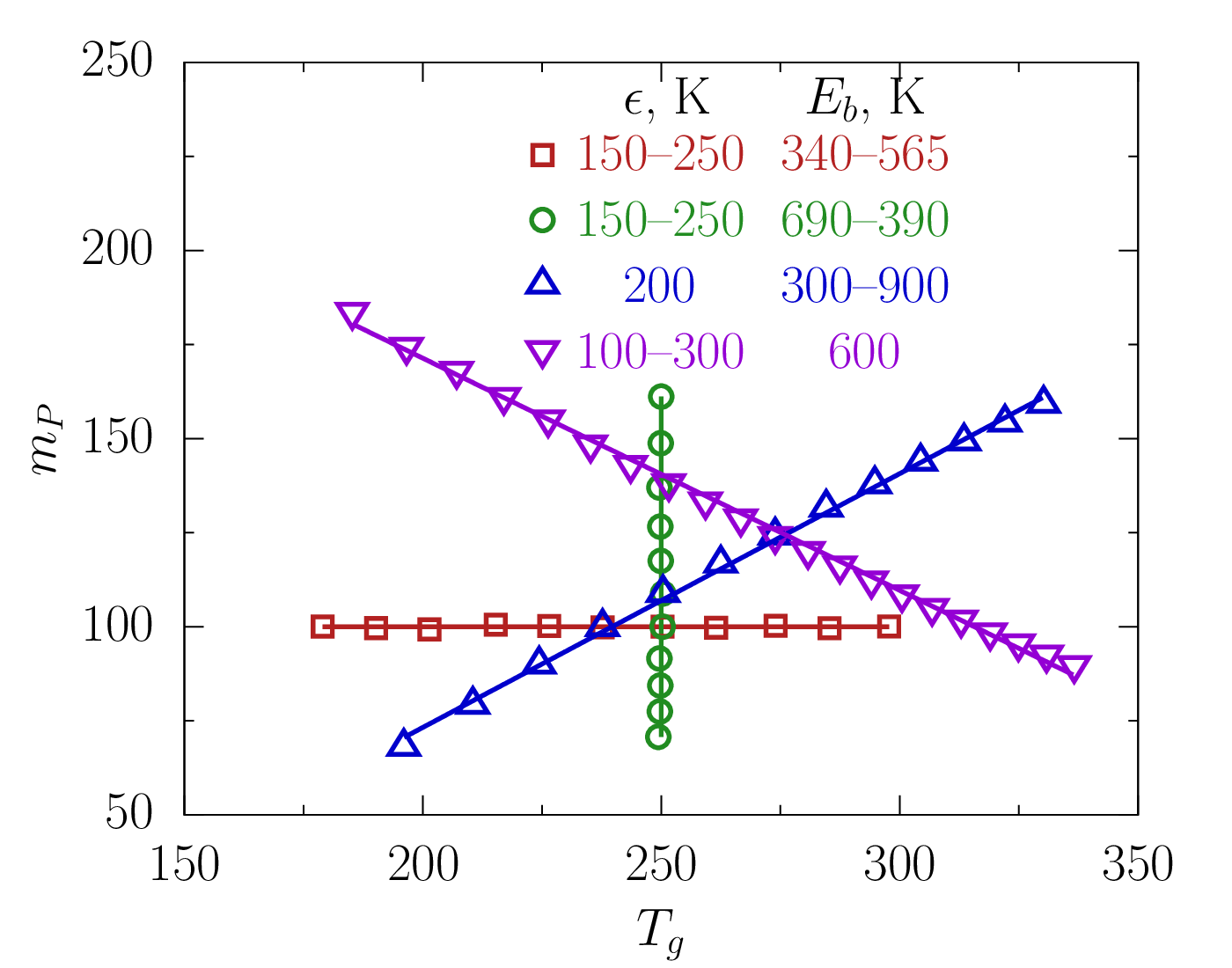}
 \caption{Diverse correlation patterns between $m_P$ and $T_g$ due to the variations in $\epsilon$ and $E_b$. The lines are a guide to the eye.}
\end{figure}

Recent experiments by Sokolov and coworkers~\cite{Mac_45_8430} show that modifying the chemical structures in the backbone/side groups of polymers produces apparently different patterns for the correlations between $m_P$ and $T_g$. For instance, $m_P$ and $T_g$ shift from $137$ and $263$ K for PP to $189$ and $304$ K for poly(vinyl alcohol) (PVA)~\cite{Mac_45_8430}, i.e., $m_P$ and $T_g$ can both increase together for some polymers. Likewise, experimental data demonstrate that $m_P$ can also decrease when $T_g$ grows; e.g., $T_g$ increases but $m_P$ decreases when shifting from poly($4$-methylstyrene) (P4MS) to poly($4$-chlorostyrene) (P4ClS). More interestingly, different polymers with similar $m_P$ (or $T_g$) may exhibit large variations in $T_g$ (or $m_P$). For instance, PVA and poly(vinyl chloride) (PVC) exhibit very similar fragilities ($m_P\approx190$ for both polymers with similar molecular weights), while their glass transition temperatures are quite different ($T_g=304$ K for PVA vs. $T_g=352$ K for PVC)~\cite{Mac_45_8430}. Hence, the variation of $m_P$ (or $T_g$) can be also nearly independent of $T_g$ (or $m_P$). Our calculations indeed support the contention that the patterns of joint variation of $m_P$ and $T_g$ can be controlled by adjusting the cohesive energy or the chain stiffness, as illustrated in Figure 7, where four correlation patterns between $m_P$ and $T_g$ are depicted, including an iso-fragility line for $m_P=100$ (where $E_b$ increases from $340$ K to $565$ K when $\epsilon$ elevates from $150$ K to $250$ K), an iso-$T_g$ line for $T_g=250$ K (where $E_b$ decreases from $690$ K to $390$ K when $\epsilon$ elevates from $150$ K to $250$ K), a positive correlation (where $E_b$ increases from $300$ K to $900$ K when the cohesive energy is fixed to be $\epsilon=200$ K), and a negative correlation (where $\epsilon$ increases from $100$ K to $300$ K when the bending energy is fixed to be $E_b=600$ K). Figure 7 exhibits the large variations in $T_g$ (or $m_P$) along the iso-fragility (or iso-$T_g$) line. Moreover, it is clear in Figure 7 that $m_P$ can increase or decrease with $T_g$ within the GET when only a single variable $E_b$ or $\epsilon$ is altered, as first revealed in ref~\citenum{JCP_131_114905}. Therefore, the GET provides a theoretical interpretation for the experimental observations in ref~\citenum{Mac_45_8430}. 

Although the concepts of fragility and glass transition temperature have long appeared in the study of glass formation~\cite{JNCS_131_13}, quantitatively understanding their molecular origins and their physical significance remains incomplete for glass-forming polymers. Based on extensive examination of the influence of various molecular factors on the properties associated with polymer glass formation, the GET suggests that the packing efficiency determines the fragility and the glass transition temperature of polymer fluids~\cite{ACP_137_125}. Our identification of iso-fragility and iso-$T_g$ lines provides additional routes for uncovering the molecular significance of fragility and the glass transition temperature by exploring the variation of typical thermodynamic properties along these lines.

\subsection{Properties along the iso-fragility and iso-$T_g$ lines}

\begin{figure}
 \centering
 \includegraphics[angle=0,width=0.45\textwidth]{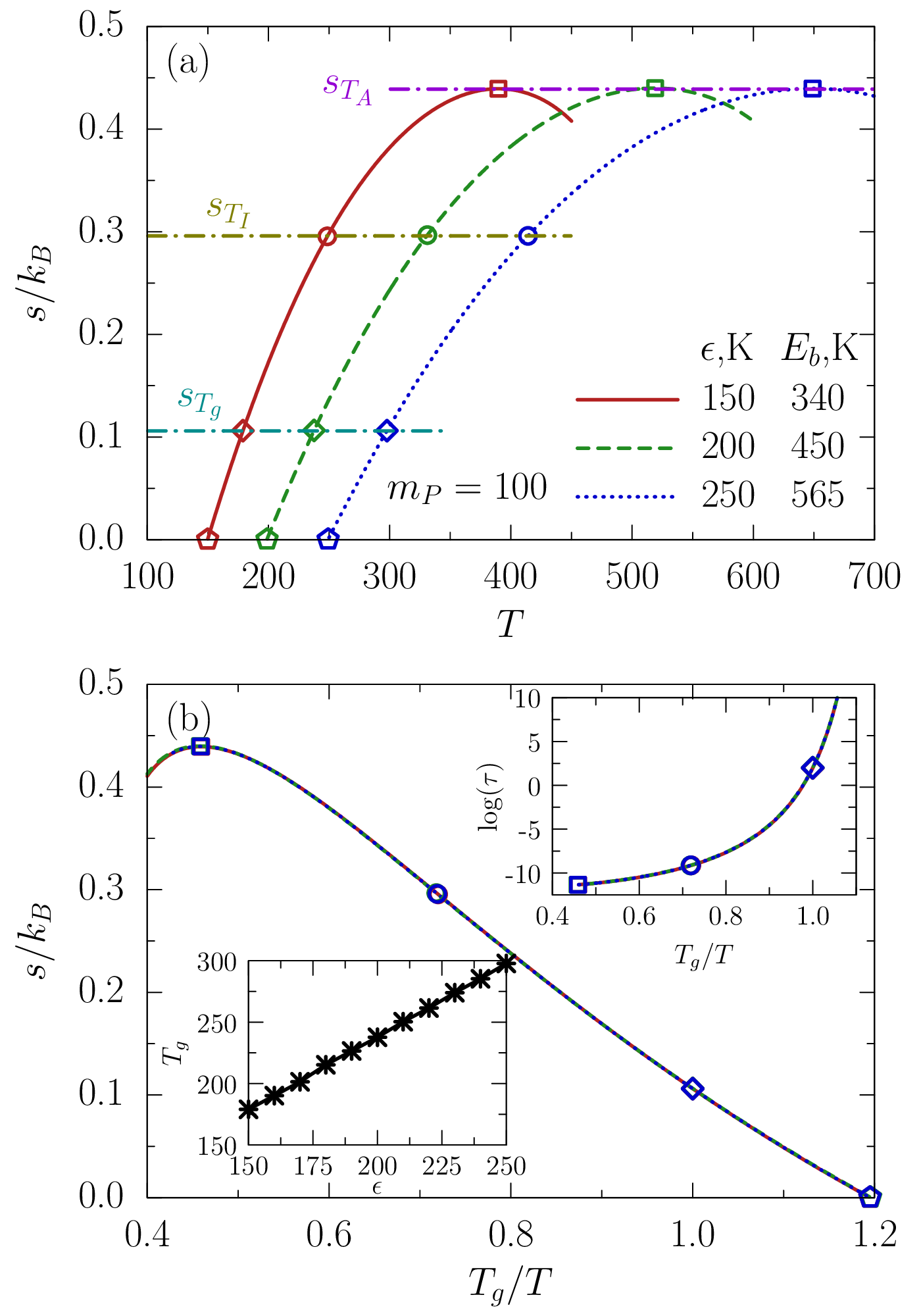}
 \caption{(a) The entropy density $s/k_B$ as a function of $T$ for several pairs of $\epsilon$ and $E_b$ that produce the same isobaric fragility parameter of $m_P=100$. (b) $T_g$-scaled Arrhenius plot for the entropy density $s/k_B$ for the same pairs of $\epsilon$ and $E_b$ as in (a). The upper inset to (b) presents the $T_g$-scaled Arrhenius plot for the relaxation time $\tau$ for the same pairs of $\epsilon$ and $E_b$ as in (a), while the lower inset to (b) depicts $\epsilon$-dependence of $T_g$ along the iso-fragility line for $m_P=100$. Squares, circles, diamonds and pentagons designate the positions of characteristic temperatures $T_A$, $T_I$, $T_g$ and $T_o$, respectively.}
\end{figure}

\begin{figure}
 \centering
 \includegraphics[angle=0,width=0.45\textwidth]{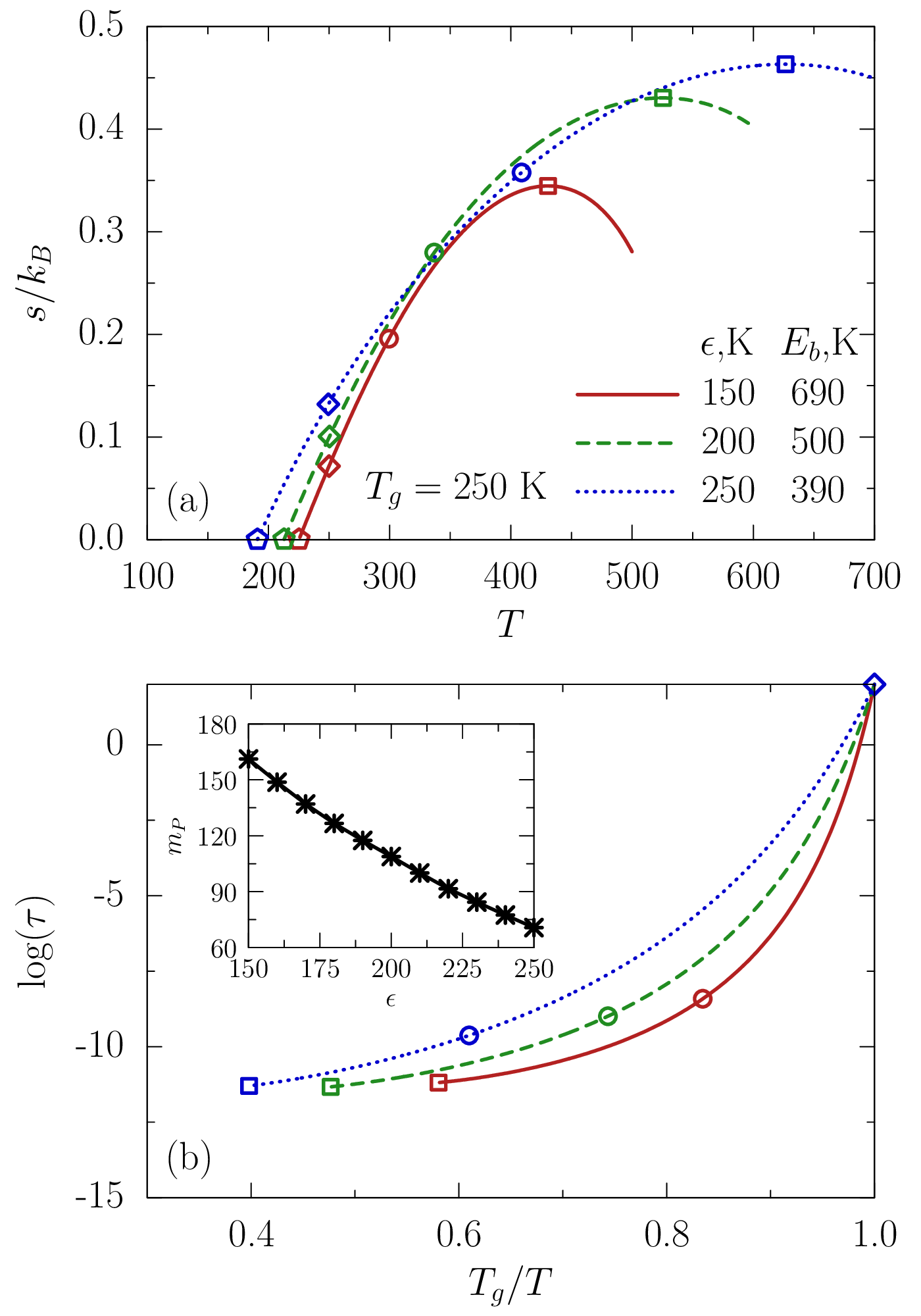}
 \caption{(a) The entropy density $s/k_B$ as a function of $T$ for several pairs of $\epsilon$ and $E_b$ that produce the same glass transition temperature of $T_g=250$ K. (b) $T_g$-scaled Arrhenius plot for the relaxation time $\tau$ for the same pairs of $\epsilon$ and $E_b$ as in (a). The inset to (b) depicts $\epsilon$-dependence of $m_P$ along the iso-$T_g$ line for $T_g=250$ K. Squares, circles, diamonds and pentagons designate the positions of characteristic temperatures $T_A$, $T_I$, $T_g$ and $T_o$, respectively.}
\end{figure}

Figure 8a displays the $T$-dependence of the entropy density for different pairs of cohesive energies $\epsilon$ and bending energies $E_b$ that lie along an iso-fragility line for $m_P=100$. As expected, the overall breadth of the entropy density curve (as measured by the characteristic temperature ratio $T_A/T_o$) remains almost unchanged along the iso-fragility lines because the fragility directly provides a measure of the breadth of glass formation, as discussed in Subsection 3.1 (Figure 5). In addition, the magnitudes of the entropy density at each characteristic temperature remain constant along the iso-fragility lines, as highlighted by the horizontal lines in Figure 8a. The above observations suggest that the entropy density along the iso-fragility lines is a unique function of $T_{\alpha}/T$, i.e., the inverse temperature $1/T$ scaled by one of the four characteristic temperatures $T_{\alpha}$. Figure 8b confirms these suggestions by presenting the entropy density along the iso-fragility line in an Angell plot. The upper inset to Figure 8b indicates that the relaxation times collapse onto a single curve when they are plotted as a function of $T_g/T$, a result that agrees with experiments~\cite{Mac_45_8430} and is just a consequence of iso-fragility lines. It is also apparent that the relaxation times at the characteristic temperature $T_A$, $T_I$ or $T_g$ remain nearly constant along the iso-fragility lines. By contrast, Figure 9 reveals that the scaling behavior is absent in the temperature dependence of the entropy density and the relaxation times along the iso-$T_g$ lines in accord with experimental data for the structural relaxation times. For instance, polymers with similar $T_g$ may have very different fragilities, resulting in different dependences of the relaxation time on $T_g/T$. Such an example can be provided by comparing the experimental data for the structural relaxation times of PVC and poly($3$-chlorostyrene) (P3ClS)~\cite{Mac_45_8430}. Figure 9a exhibits the noticeable elevation of the entropy density at $T_A$, $T_I$ or $T_g$ with increasing $\epsilon$ along the iso-$T_g$ lines.

\begin{figure}
 \centering
 \includegraphics[angle=0,width=0.45\textwidth]{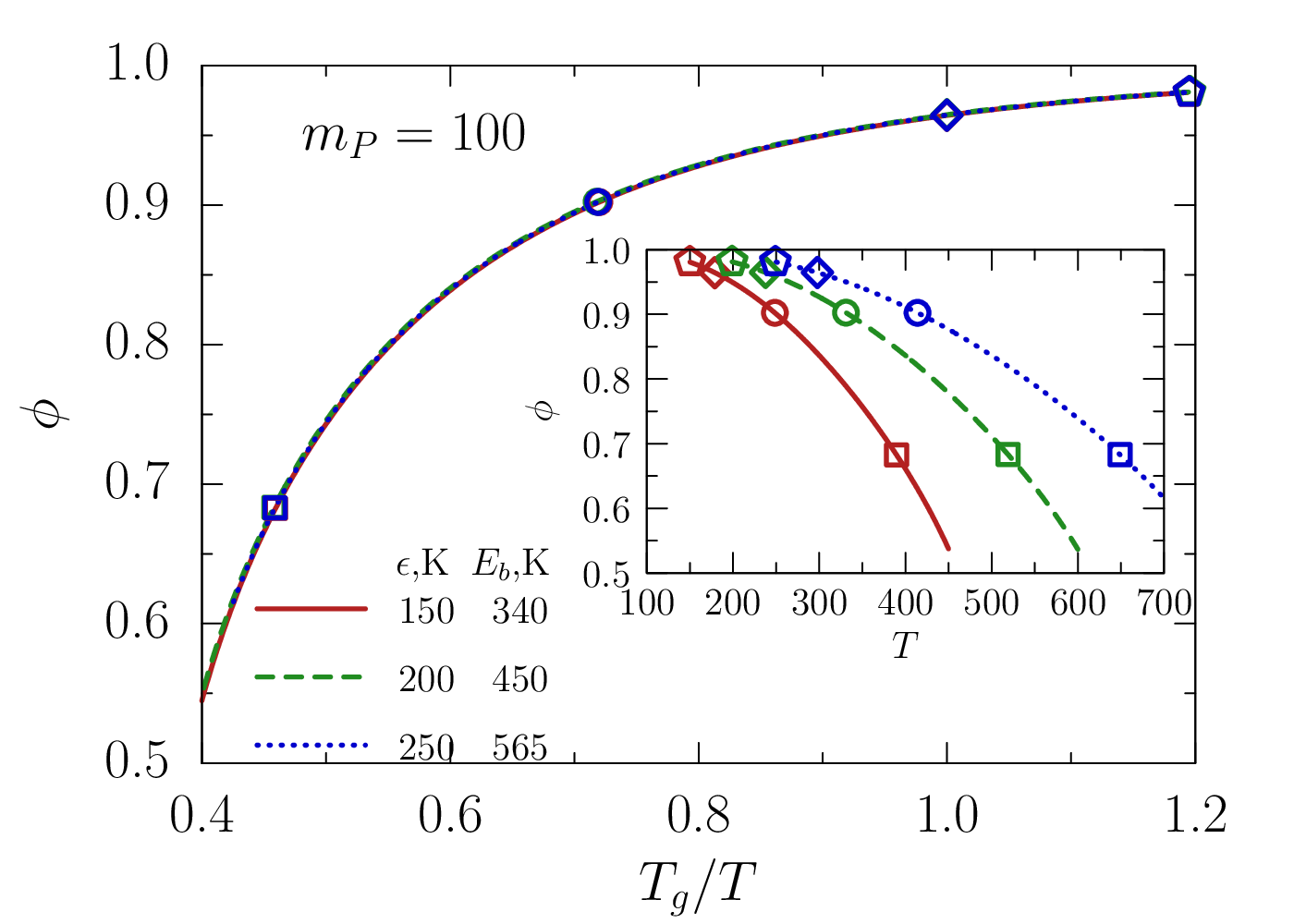}
 \caption{$T_g$-scaled Arrhenius plot of the polymer volume fraction $\phi$ for several pairs of $\epsilon$ and $E_b$ that produce the same isobaric fragility parameter of $m_P=100$. The inset depicts $T$-dependence of $\phi$. Squares, circles, diamonds and pentagons designate the positions of characteristic temperatures $T_A$, $T_I$, $T_g$ and $T_o$, respectively.} 
\end{figure}

Figure 10 further reveals that the polymer volume fraction $\phi$ along the iso-fragility lines becomes a unique function of $T_g/T$ and that the volume fraction at each characteristic temperature is independent of $\epsilon$ and $E_b$. Again, such scaling is absent from the EOS along the iso-$T_g$ lines (data not shown). Instead, our computations indicate that the polymer volume fraction at each characteristic temperature increases with $\epsilon$ along the iso-$T_g$ lines (see Figure S2 in the Supporting Information), a trend that can be explained by the negative correlation between $\epsilon$ and $E_b$ along the iso-$T_g$ lines. The temperature dependence of the polymer volume fraction, of course, changes significantly along both iso-fragility (see the inset to Figure 10) and iso-$T_g$ lines due to the variations in the cohesive energy. 

Figures 8 and 9 exhibit the characteristic temperatures as depending differently on $\epsilon$ along the iso-fragility and iso-$T_g$ lines, respectively. The general increase of $T_{\alpha}$ with $\epsilon$ along iso-fragility lines arises (Figure 9a) because the bending energy $E_b$ increases with $\epsilon$ along the iso-fragility lines. The simultaneous increase of $\epsilon$ and $E_b$ inevitably leads to the elevation of all characteristic temperatures (see Figure 2). Moreover, all characteristic temperatures grow linearly with $\epsilon$ along the iso-fragility lines within the parameter range investigated (see the lower inset to Figure 8b and more results in Figure S3 in the Supporting Information). The dependence of the characteristic temperatures on $\epsilon$ appears to be complicated along the iso-$T_g$ lines. For example, $T_A$ and $T_I$ are found to increase with $\epsilon$ (Figures S4a and S4b in the Supporting Information), while $T_o$ undergoes a slight drop with $\epsilon$ (Figure S4c in the Supporting Information). On the other hand, the fragility parameter $m_P$ monotonically diminishes as a function of $\epsilon$ along the iso-$T_g$ lines (see the inset to Figure 9b and more results in Figure S4d), a trend explained by the GET since $E_b$ decreases with $\epsilon$ along the iso-$T_g$ lines and since $m_P$ diminishes as either $\epsilon$ increases or $E_b$ decreases.

\subsection{Implications of the influence of cohesive energy and chain stiffness on polymer glass formation}

Although the general variations along the iso-fragility and iso-$T_g$ lines of all the properties considered in Subsection III C can be explained by analyzing the combined influence of the cohesive and bending energies, the scaling properties displayed by the entropy density and by the EOS along the iso-fragility lines are first derived here from the GET. Those remarkable scaling properties along the iso-fragility lines and their absence along the iso-$T_g$ lines support the well-known contention that fragility provides more fundamental insight into glass formation than the somewhat arbitrarily defined glass transition temperature $T_g$. For instance, polymer fragility has been correlated with a variety of properties of glass formation~\cite{JCP_114_5621, JCP_99_4201, Mac_24_1222, Science_267_1945, PRL_71_2062, Science_273_1675}. By contrast, $T_g$ is often defined empirically and depends on the cooling rate for some types of experiments. Nevertheless, $T_g$ is still of great importance in characterizing glass formation because certain properties exhibit special features around $T_g$. Moreover, $T_g$ is a crucial parameter governing practical applications of glassy materials since $T_g$ signals the presence of drastic changes in the mechanical and rheological properties of the materials~\cite{Science_267_1924}. Our calculations suggest that controlling the cohesive energy and chain stiffness enables finely tailoring the fragility and the glass transition temperature of glass-forming polymers and hence provides guidance towards the rational design of polymer materials.

\section{Summary}

We examine the influence of cohesive energy and chain stiffness on polymer glass formation using the generalized entropy theory in conjunction with a minimal model of polymer melts with monomer averaged interactions, where a single nearest neighbor van der Waals energy is employed to describe the interactions between all pairs of nearest neighbor united atom groups. Polymers with rigid structures and weak nearest neighbor interactions are demonstrated as being the most fragile, while the glass transition temperature $T_g$ becomes greater when the chains are stiffer and/or the nearest neighbor interactions are stronger. We find two simple algebraic expressions for describing the calculated dependence of the isobaric fragility parameter $m_P$ and $T_g$ on $\epsilon$ and $E_b$.
 
The strong influence of the cohesive and bending energies naturally inspires the introduction of some important design concepts, such as iso-fragility and iso-$T_g$ lines. Analysis of relevant properties along these special lines provides further evidence that fragility plays a more fundamental role in the description of glass formation than $T_g$. The present work clearly implies that controlling the cohesive energy and chain stiffness enables finely tailoring the fragility and the glass transition temperature of glass-forming polymers and hence provides an efficient route for guiding the rational design of polymer materials. Finally, we note that the results presented here will be compared with those from more detailed models of polymer melts that have specific interactions for particular united atom groups to study and extend the limits of validity of the minimal model of melts with monomer averaged interactions.

\begin{acknowledgments}
We thank Jack Douglas for helpful discussions and a critical reading of the manuscript. This work is supported by the U.S. Department of Energy, Office of Basic Energy Sciences, Division of Materials Sciences and Engineering under Award DE-SC0008631.
\end{acknowledgments}

\bibliography{aps}% Produces the bibliography via BibTeX.

%%%%%%%%%% Merge with supplemental materials %%%%%%%%%%
\pagebreak
\widetext
\clearpage
\begin{center}
\textbf{\large Supporting Information}
\end{center}
%%%%%%%%%% Merge with supplemental materials %%%%%%%%%%
%%%%%%%%%% Prefix a "S" to all equations, figures, tables and reset the counter %%%%%%%%%%
\setcounter{equation}{0}
\setcounter{figure}{0}
\setcounter{table}{0}
\setcounter{page}{1}
\makeatletter
\renewcommand{\theequation}{S\arabic{equation}}
\renewcommand{\thefigure}{S\arabic{figure}}
\renewcommand{\bibnumfmt}[1]{[S#1]}
\renewcommand{\citenumfont}[1]{S#1}
%%%%%%%%%% Prefix a "S" to all equations, figures, tables and reset the counter %%%%%%%%%%

\begin{figure}[b]
 \centering
 \includegraphics[angle=0,width=0.75\textwidth]{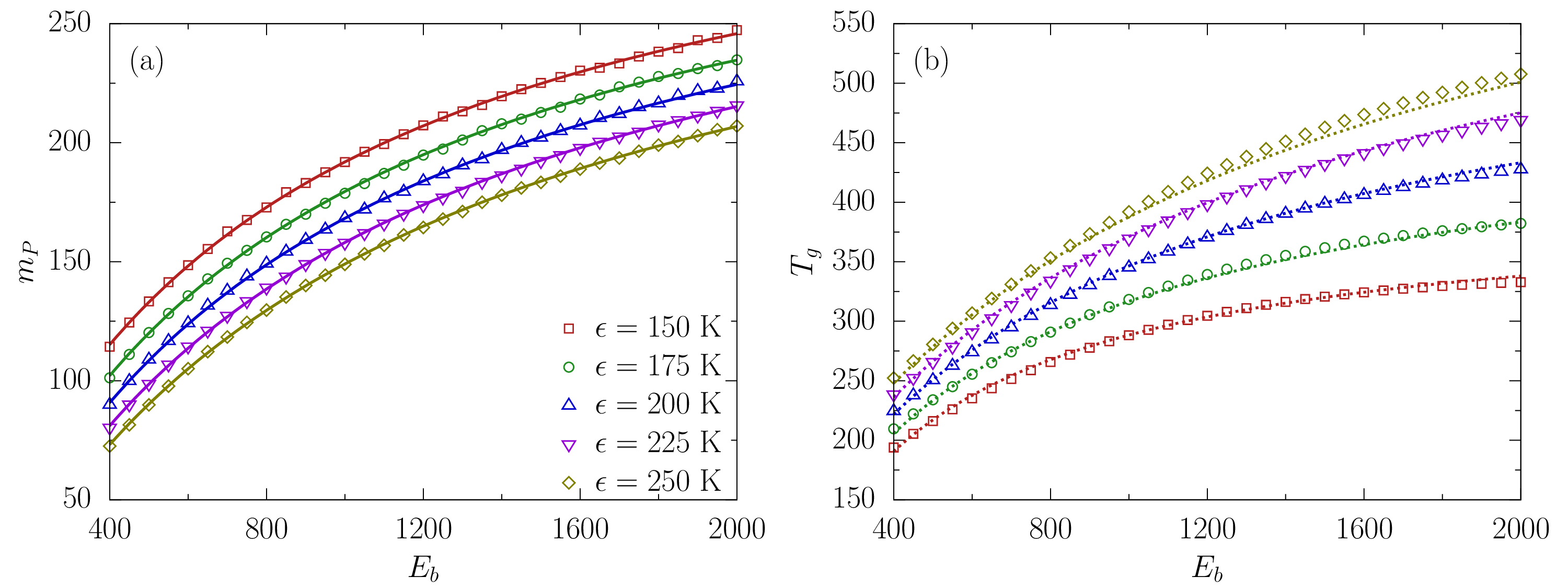}
 \caption{(a) The isobaric fragility parameter $m_P$ and (b) the glass transition temperature $T_g$ as a function of the bending energy $E_b$ for various cohesive energies $\epsilon$. Solid lines in (a) and dotted lines in (a) are fits according to eqs 4 and 5 in the main text with the fitting parameters $a_0=1.23154\times10^2$, $a_1=-1.17539$, $a_2=2.23509\times10^{-2}$, $b_0=0.606943$, $b_1=-1.07595\times10^{-3}$, $c_0=1.49847\times10^{-3}$, $c_1=-5.60474\times10^{-7}$, $c_2=-4.78656\times10^{-9}$ and $u_0=37.8879$, $u_1=50325$, $u_2=-1.0445\times10^{7}$, $v_0=1.80486$, $v_1=-609.037$, $v_2=8.08886\times10^{4}$, $w_0=4.7747\times10^{-3}$, $w_1=-2.09936$, $w_2=282.37$. As in the main text, the following parameters are used here and in the subsequent figures: the lattice coordination number is $z=6$, the pressure is $P=1$ atm, the cell volume parameter is $a_{\text{cell}}=2.7$\AA{}, and the polymerization index is  $N_c=8000$.}
\end{figure}

\textbf {S1 Fitting results for $m_P(\epsilon, E_b)$ and $T_g(\epsilon, E_b)$}---In the main text, we propose two simple algebraic equations that fairly accurately capture the computed combined variations of the isobaric fragility parameter $m_P$ and the glass transition temperature $T_g$ with the microscopic cohesive $\epsilon$ and bending $E_b$ energies. Figure S1 displays the $E_b$-dependence of $m_P$ and $T_g$ for various $\epsilon$, along with our best fits obtained from eqs 4 and 5 in the main text. The fitting parameters are summarized in the caption of Figure S1. 

\begin{figure}[b]
 \centering
 \includegraphics[angle=0,width=0.6\textwidth]{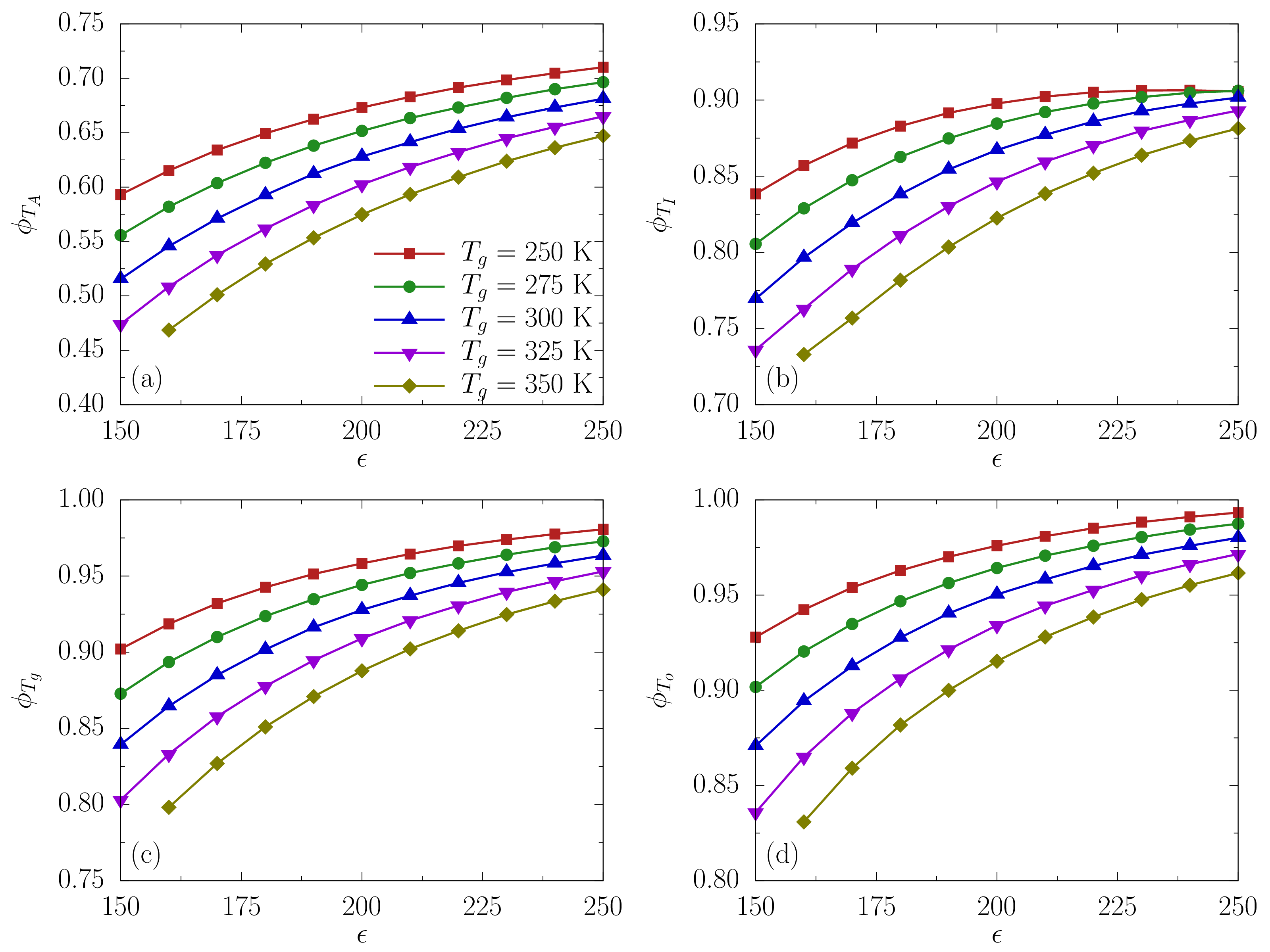}
 \caption{The volume fractions at different characteristic temperatures $\phi_{T_{\alpha}}$ as a function of $\epsilon$ along selected iso-$T_g$ lines with the indicated values of $T_g$. (a) $\phi_{T_A}$. (b) $\phi_{T_I}$. (c) $\phi_{T_g}$. (d) $\phi_{T_o}$.}
\end{figure}

\textbf {S2 More properties along the iso-fragility and iso-$T_g$ lines}---Figure 10 in the main text reveals that the polymer volume fraction $\phi$ along the iso-fragility lines becomes a unique function of $T_g/T$ and that the volume fraction at each characteristic temperature is independent of $\epsilon$ and $E_b$. By contrast, this behavior is not observed along the iso-$T_g$ lines. Figure S2 indicates that the polymer volume fraction at each characteristic temperature increases with $\epsilon$ along the iso-$T_g$ lines, a trend that can be explained by the negative correlation between $\epsilon$ and $E_b$ along the iso-$T_g$ lines (see Figure 6b in the main text).

\begin{figure}
 \centering
 \includegraphics[angle=0,width=0.6\textwidth]{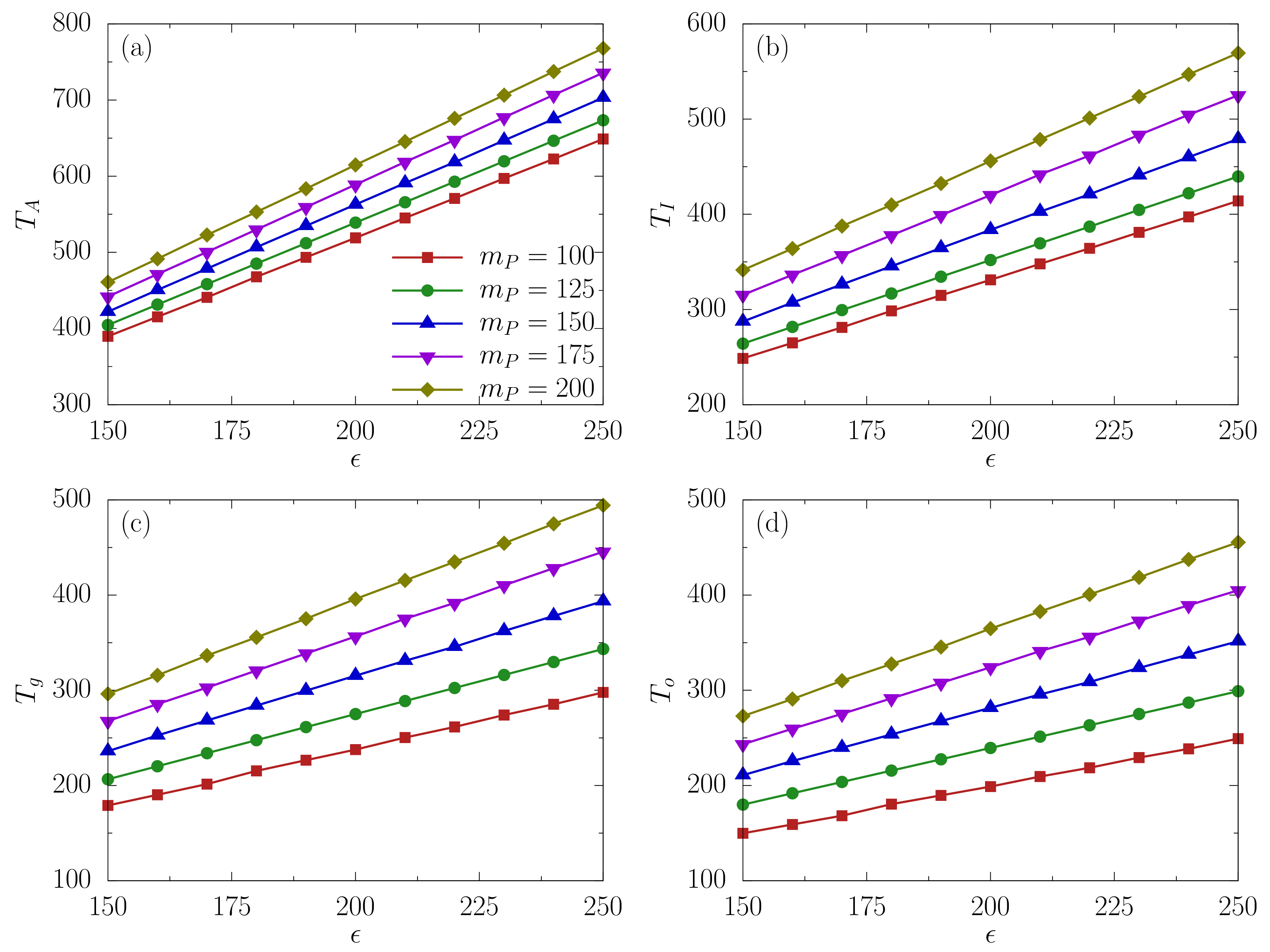}
 \caption{Characteristic temperatures as a function of $\epsilon$ along selected iso-fragility lines with the indicated values of $m_P$. (a) $T_A$. (b) $T_I$. (c) $T_g$. (d) $T_o$.}
\end{figure}

The lower inset to Figure 8b in the main text reveals that $T_g$ grows linearly with $\epsilon$ along an iso-fragility line for $m_P=100$. Figure S3 indicates that the linear relationship also holds for other characteristic temperatures and other $m_P$.

\begin{figure}
 \centering
 \includegraphics[angle=0,width=0.6\textwidth]{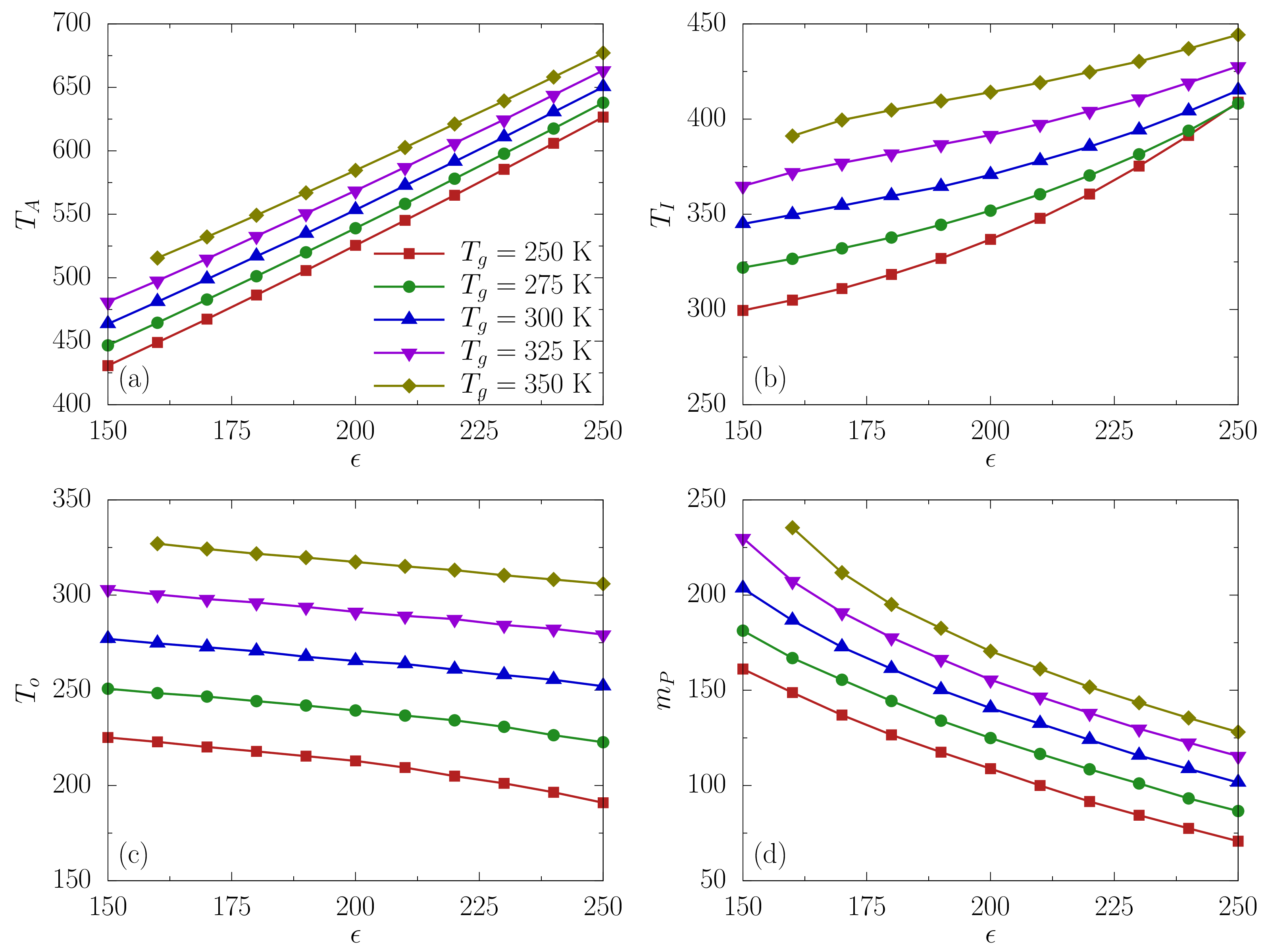}
 \caption{Characteristic temperatures and fragility parameter as a function of $\epsilon$ along selected iso-$T_g$ lines with the indicated values of $T_g$. (a) $T_A$. (b) $T_I$. (c) $T_g$. (d) $m_P$.}
\end{figure}

Figure S4 displays the $\epsilon$-dependence of the characteristic temperatures and fragility parameter along the iso-$T_g$ lines with representative $T_g$. Both $T_A$ and $T_I$ increase with $\epsilon$ (Figures S4a and S4b), while $T_o$ undergoes a slight drop with $\epsilon$ (Figure S4c). On the other hand, the fragility parameter $m_P$ monotonically diminishes as a function of $\epsilon$ along the iso-$T_g$ lines (Figure S4d), a trend explained in the main text since $E_b$ decreases with $\epsilon$ along the iso-$T_g$ lines (Figure 6b in the main text) and since $m_P$ diminishes as either $\epsilon$ increases or $E_b$ decreases (Figure 4 in the main text).

%\begin{thebibliography}{11}
%\bibitem{S_RefA} A. Someone, C. Someone, D. Someone, Phys. Rev. Lett. {\bf 11}, 1111 %(1911).
%\end{thebibliography}

\end{document}